\documentstyle[prb,aps,epsf,twocolumn]{revtex} 
\begin{document}

\title{Long range magnetic ordering in a spin-chain compound, Ca$_3$CuMnO$_6$, with multiple bond distances}

\author{Kausik Sengupta, S. Rayaprol,  Kartik K Iyer, and E.V. Sampathkumaran \footnote{Corresponding author: Email address: sampath$@$tifr.res.in}}

\address{Tata Institute of Fundamental Research, Homi Bhabha Road, 
Mumbai - 400 005, INDIA.}

\maketitle

\begin{abstract} 
{The results of ac and dc magnetization and heat capacity measurements as a function of temperature (T = 1.8 to 300 K) are reported for a quasi-one-dimensional compound, Ca$_3$CuMnO$_6$, crystallizing in a triclinically distorted K$_4$CdCl$_6$-type structure. The results reveal  that this compound undergoes  antiferromagnetic ordering close to 5.5 K. In addition, there is another magnetic transition below 3.6 K. Existence of two  long-range magnetic transitions is uncommon among quasi-one-dimensional systems. It is interesting to note that both the magnetic transitions are of long-range type,  instead of spin-glass type, in spite of the  fact that the Cu-O and Mn-O bond distances are multiplied due to this crystallographic  distortion.  In view of this, this compound could serve as a nice example for studying "order-in-disorder" phenomena.   }
\end{abstract}
\vskip0.5cm
$^*$E-mail address: sampath@tifr.res.in
\vskip1cm

\maketitle
The spin-chain compounds of the type, (Sr, Ca)$_3$MXO$_6$ (M, X= a metallic ion, magnetic or nonmgnetic), crystallizing in the K$_4$CdCl$_6$ (rhomhohedral) derived structure (space group $R{\bar{3}}c$), are attracting a lot   attention in the recent literature (see, for instance, Refs. 1-17 and references cited therein). The structure is characterized by the presence of chains of M and X ions running along c-direction arranged hexagonally forming a triangular lattice. These chains are  separated by Sr (or Ca) ions. Within the chains, alternating MO$_6$ trigonal prism and XO$_6$ octahedra share one of the  faces.  The triangular  arrangement   of magnetic ions in the a-b plane may result in magnetic frustration in the event that the interchain interaction on the a-b plane is antiferromagnetic. A survey of the  literature reveals that a variety of magnetic behaviour, including "partially disordered antiferromagnetic structure" (a rare magnetic structure), has been observed (Refs. 10, 11) in this class of compounds. Among these spin-chain oxides, the Cu containing ones are especially of interest due to a lowering of crystal symmetry. For instance, Sr$_3$CuIrO$_6$ undergoes a monoclinic distortion (space group C2/c) due to Jahn-Teller effect as a result of which the Cu ions at the trigonal prisms move slightly away within the basal plane from its original position thereby attaining pseudo-square-planar oxygen coordination.\cite{1} An interplay of this distortion and defects makes the magnetism inhomogeneous\cite{11} with the magnetic behavior being sensitive to sample preparative conditions.\cite{12,13}  One would therefore expect that such distortions should not result in long range magnetic ordering. Therefore, we carried out magnetic studies on another Cu containing compound, Ca$_3$CuMnO$_6$, which has been recently synthesized\cite{17} and found to undergo further distortion to a triclinic structure (space group, P-1). The results reveal that this compound actually undergoes antiferromagnetic ordering close to 5.6 K, in contrast to the inhomogeneous, spin-glass behaviour seen for Sr$_3$CuIrO$_6$, thereby presenting another interesting situation among this class of compounds. In addition, interestingly, there is one more magnetic transition below 3.6 K, a feature uncommon among other classes of quasi-low-dimensional magnetic materials. 

The title compound was prepared by a conventional solid state route from high purity ($> 99.9\%$) CaCO$_3$, MnO$_2$ and CuO. Required amounts of these were thoroughly mixed, pressed into pellets and annealed at 1130 C for 48 h.  The pellets were ground finely, pelletized and heat-treated in an atmosphere of oxygen at 1000 C for 4h, after which it was slow-cooled. The samples were characterized by x-ray diffraction and the  pattern (see Fig. 1) confirms single phase nature of the specimen. The dc magnetic measurements were performed by commercial superconducting quantum interference device (Quantum 
Design) and vibrating sample (Oxford Instruments) 
\begin{figure}
\centerline{\epsfxsize=7.5cm{\epsffile{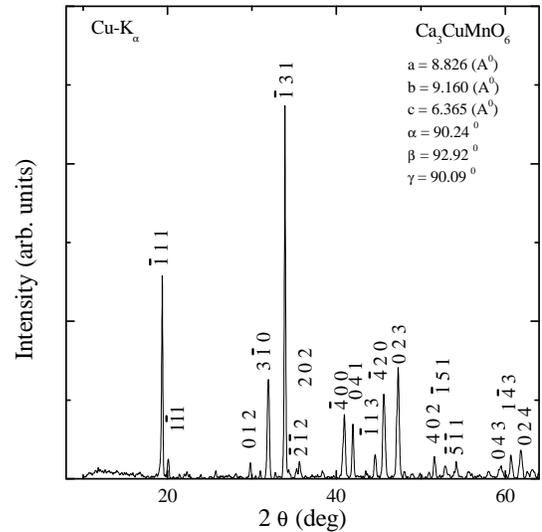}}}
\caption{X-ray diffraction pattern (Cu K$_\alpha$) of Ca$_3$CuMnO$_6$ at room temperature. Intense peaks are indexed. The lattice parameters and Miller indices are also included in the figure.}
\end{figure}
\begin{figure}
\centerline{\epsfxsize=7.5cm{\epsffile{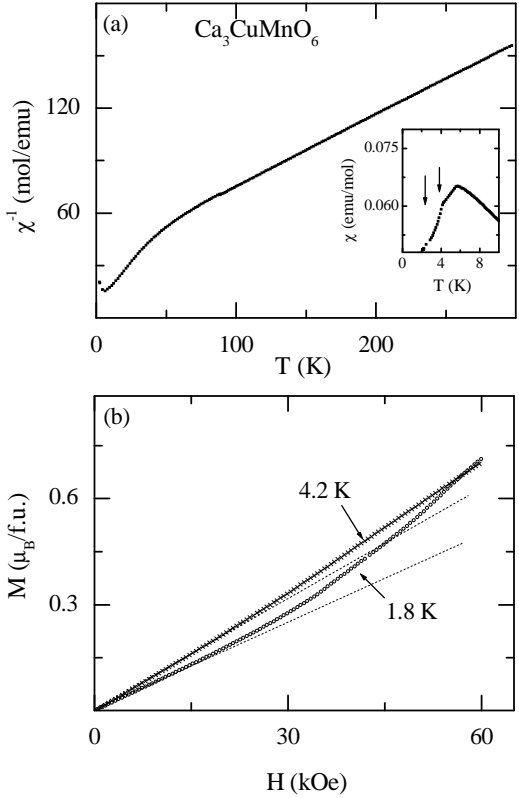}}}
\caption{(a) Inverse susceptibility ($\chi$) as a function of temperature measured in a magnetic field of 5 kOe for Ca$_3$CuMnO$_6$.  Inset shows the low temperature $\chi$ data in an expanded form to highlight the features due to magnetic ordering. Vertical arrows mark the two transitions. (b) Isothermal magnetization at selected temperatures and the linear lines extrapolated from low field data are also shown.}
\end{figure}
magnetometers, whereas ac susceptibility ($\chi$) data were obtained with the former. Heat-capacity (C) data (1.8 - 40 K) were obtained by a semi-adiabatic heat-pulse method. 

The results of dc $\chi$ measurements recorded in the presence of a magnetic field (H) of 5 kOe is shown in Fig. 2a. The $\chi (T)$ follows Curie-Weiss behaviour above 50 K. The paramagnetic Curie temperature $(\theta_{p})$ obtained from this plot turns out to be about -85 K with the negative sign indicating the existence of antiferromagnetic correlations. The effective moment  is found to be 4.4$\mu_B$, which is close to the value obtained assuming Cu and Mn are 2+ and 4+ states respectively.  As the temperature is lowered below 50 K, there is a deviation of inverse $\chi (T)$ curve from  high temperature linear behaviour, presumably due to short range magnetic correlations, leading to magnetic ordering close to  5.7 K  as evidenced by the appearance of a distinct peak at 5 K. There is an additional shoulder at 3.6 K as though there is another magnetic transition. In order to understand the nature of the magnetic ordering, we have performed $\chi$ measurements  at low temperatures (2-20 K) in the presence of a H of 100 Oe, both for field-cooled (FC) and zero-field-cooled (ZFC) conditions of the specimen, as well as isothermal magnetization (M) measurements. It is clear from the figure 3a that there is a distinct peak at 5.7 K. The  $\chi$(T) plots for both ZFC and FC states merge and this is sufficient enough to establish that both the magnetic transitions are of a long-range type, presumably of an antiferromagnetic type judged by the appearance of a peak in the $\chi (T)$ plot at the onset of magnetic ordering. Further evidence for this conclusion is obtained from the data shown in Fig. 2b. At 1.8 K, M varies linearly with H till about 30 kOe, which is followed by a  curvature of the plot of M(H) at higher fields to a higher M value compared to the one obtained from the linear extrapolation of the low field data as though there is a spin reorientation. There is no evidence for saturation till the highest field applied  and the plot is non-hysteretic. Similar features are seen at 4.2 K, but not at 10 K (not shown in the figure). These features establish that both the transitions are of an antiferromagnetic type. It is at present not clear whether both Cu and Mn ions contribute to these magnetic transitions.   

\begin{figure}
\centerline{\epsfxsize=7.5cm{\epsffile{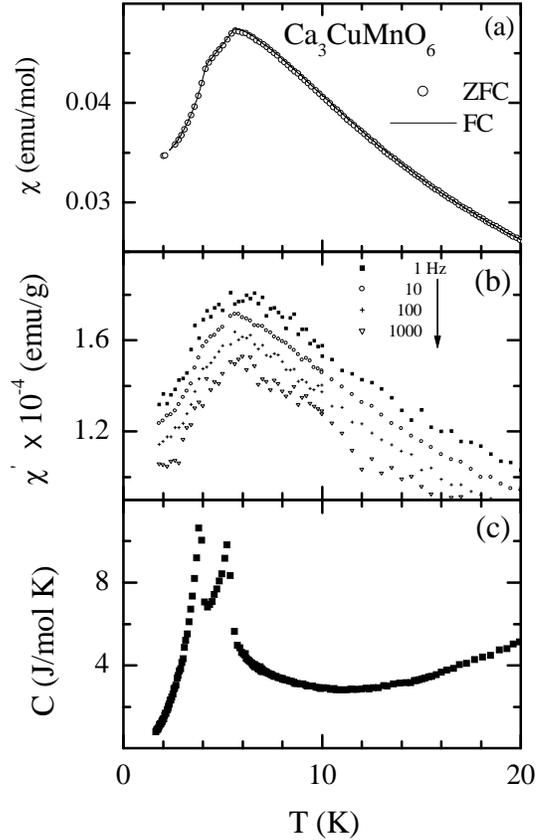}}}
\caption{(a) Low temperature magnetic susceptibility ($\chi$) behaviour measured in 100 Oe for Ca$_3$CuMnO$_6$. Zero-field-cooled data are shown by points, whereas the continuous line represents the data for the field-cooled state.  (b) Real part of ac $\chi$ measured at various frequencies. The curves  for all the frequencies overlap and hence are shifted downwards with increasing frequency for the sake of clarity. (c) Heat-capacity as a function of temperature.}  
\end{figure}

Further evidence for the absence of spin-glass freezing is obtained from the ac $\chi$ data (1.8 - 20 K), recorded  at various frequencies ($\nu$= 1, 10, 100 and 1000 Hz). In the real part ($\chi^{\prime}$) of ac $\chi$, there is a broad peak close to 5.7 K, in addition to a weak shoulder at about 3.6 K, as clearly seen in the data for $\nu$= 10 Hz (in which case the data is less noisy), consistent with the two magnetic transitions inferred above (Fig. 3b). It is to be noted that the peak temperature is not frequency-dependent. In addition, the imaginary  part of ac $\chi$ is completely featureless in the temperature range of investigation (and hence not shown in the form of a figure). These findings conclusively establish that both the magnetic orderings are not of a spin-glass type. We have also measured C as a function of temperature (Fig. 3c). There are two prominent peaks in C(T) plot, one at 3.7 K and the other at 5.2 K; the huge jumps at the transitions establish that both the transitions are of bulk nature thereby ruling out a role of  magnetic impurities (not detectable by x-ray diffraction) on the origin of either of the two transitions.  The sharpness of both the transitions in this plot is consistent with long range magnetic order, but not with random freezing of the spins. If the spins undergo spin-glass freezing, then the  peak in C should be broader without $\lambda$-anomaly; in addition, the peak in C should appear\cite{18} at a temperature which is higher by about 20$\%$ compared to the transition  temperature, given by, say the peak temperature in ac $\chi$, and C should vary linearly with T well below magnetic transition temperature, in sharp contrast to the observations. Therefore, spin-glass freezing is completely ruled out.  

Thus, the results presented above conclusively establish that this compound exhibits antiferromagnetic ordering close to (T$_N$) 5.7 K. A careful look at the crystallographic features\cite{17}  indicates that this finding is intriguing. In contrast to Sr$_3$CuIrO$_6$, there is a partial replacement of Cu sites (to the extent of 10$\%$) by X ions (in this case Mn), however maintaining the characteristic coordination (4 + 2)  for all the Cu ions.  As a result of this atomic disorder, Jahn-Teller effect is partially suppressed for this fraction of Cu, which means that there is an increase of Cu-O mean distances for 4-coordination, though still all the six Cu-O bond distances are totally different. This means that the positions of these Cu ions are more towards the center of the trigonal prism, compared to the  rest of Cu ions, which remain closer to a face of trigonal prism (that is, pseudo square-planar coordination) as in the case of Sr$_3$CuIrO$_6$.  The mean 4-coordination Cu-O distances for the latter case are  thus reduced to 2.026 \AA $~$  compared to that (2.122 \AA)   of   former Cu ions.  In the same way, there are multiple Ca-O and Cu-Mn distances.      
Thus, in spite of randomness of Cu-O and Mn-O bond distances and Cu-Mn disorder, it is fascinating that this compound does not exhibit a corresponding frustration in the magnetic structure. The only role of frustration presumably is to reduce the magnetic ordering temperature, considering that the value of $\theta_{p}$ is comparatively larger. Therefore, the observation of long range magnetic ordering implies that the site-exchange takes place in an ordered fashion. Thus, this compound serves as a nice example for "order in disorder" phenomena. In this respect, it is worthwhile to carry out crystal structure studies at low temperatures  in order to ensure that there is no change in crystal symmetry in the vicinity of magnetic ordering temperature.



\end{document}